\documentstyle[aps,prl,psfig,multicol]{revtex}

\newcommand{\beq}{\begin{equation}}
\newcommand{\eeq}{\end{equation}}
\newcommand{\bea}{\begin{eqnarray}}
\newcommand{\eea}{\end{eqnarray}}

\begin{document}

\draft

\title{\bf Topological origin of the phase transition in a mean-field 
model}

\author{Lapo Casetti$^{1,}$\cite{lapo}, E. G. D. Cohen$^{2,}$\cite{egd} and Marco Pettini$^{3,}$\cite{marco}}

\address{$^1$Istituto Nazionale per la Fisica della Materia (INFM), Unit\`a di Ricerca del Politecnico di Torino,\\
Dipartimento di Fisica, Politecnico di Torino,
Corso Duca degli Abruzzi 24, I-10129 Torino, Italy} 

\address{$^2$The Rockefeller University, 1230 York Avenue, New York, New York 
10021 - 6399} 

\address{$^3$Osservatorio Astrofisico di Arcetri, Largo Enrico Fermi 5,
I-50125 Firenze, Italy,\\ 
and Istituto Nazionale per la Fisica della Materia (INFM), Unit\`a di Ricerca
di Firenze, Italy} 

\date{\today}

\maketitle

\begin{abstract}
We argue that the phase transition in the mean-field XY model is related
to a particular change in the topology of its configuration space. The nature of
this topological change can be discussed on the basis of elementary Morse theory 
using the potential energy per particle $\cal V$ as a Morse function. 
The value of $\cal V$ where such a topological change occurs equals the 
thermodynamic value of $\cal V$ at the phase transition and the number of 
(Morse) critical points grows very fast with the number of particles $N$. 
Furthermore, as in statistical mechanics, also in topology the way the 
thermodynamic limit is taken is crucial.
\end{abstract}

\begin{multicols}{2}
\narrowtext

It is customary in statistical mechanics to associate phase transitions 
with singularities in the equilibrium measure that describes the macroscopic 
system in its phase space. Recently it has been conjectured \cite{cccp} 
that second-order phase transitions
might have a topological origin, i.e., that a thermodynamic 
transition might be related to a change in the topology of the configuration 
space, and that the observed singularities in the statistical-mechanical equilibrium measure and in the thermodynamic
observables at the phase transition might be interpreted as a ``shadow'' of 
this major topological change that happens
at a more basic level. Such a conjecture has been put forward heuristically, 
based on numerical simulations where the averages and fluctuations
--- either time or statistical-mechanical --- for some
observables of a geometric nature (e.g. configuration-space curvature fluctuations) related to the Riemannian geometrization of the dynamics in configuration space have been computed. When plotted 
as a function of either the temperature or the energy, the fluctuations of the curvature have a singular behavior at the transition point which
can be qualitatively reproduced using a geometric model. In such a model 
the origin of the singular behavior of the curvature fluctuations resides
in a topological change \cite{cccp}. Extensive numerical work \cite{ext_papers} has shown that the curvature fluctuations indeed exhibit qualitatively the same singular behavior in many different models undergoing continuous phase transitions, namely $\varphi^4$ lattice models with discrete and continuous symmetries and XY models. 
The presence of a singularity in the statistical-mechanical averages
as well as in the fluctuations of the curvature at the transition point has 
been proved analytically for the mean-field XY-model \cite{Firpo}, 
which is the model that we will consider in the following. 
Moreover, a purely geometric, and thus still indirect, further indication 
that the topology of the configuration space might change at the
phase transition has been obtained from numerical calculations
for the $\varphi^4$ model on a two-dimensional lattice \cite{Franz}. 

In spite of all these indirect indications, no {\em direct} evidence, 
analytical nor numerical, 
of the presence of a change in the topology of the configuration 
manifold that
can be related with a thermodynamic phase transition had yet been found.
The purpose of the present Letter is to argue that such a topological change 
in configuration space indeed exists in the particular case of the mean-field 
XY model, and in doing so to clarify 
the nature of the relationship between topologic and thermodynamic transitions, discussing especially the role played by the thermodynamic limit. 

Let us now introduce the mathematical tools we need to detect and characterize topological changes, which are referred to as 
elementary critical point (Morse) theory \cite{Morse_theory}. Such a theory  links the topology of a given manifold with the properties of 
the critical points of functions defined on it. Given a (compact) 
manifold $M$ and a smooth function $f: M \mapsto {\bf R}$, a
point $x_c \in M$ is called a {\em critical point} of $f$ if $df = 0$, while
the value $f(x_c)$ is called a {\em critical value}. A level set
$f^{-1}(a) = \{ x \in M : f(x) = a \}$ of $f$ is called a {\em critical level}
if $a$ is a critical value of $f$, i.e., if there is at least one 
critical point $x_c \in f^{-1}(a)$. The function $f$ is called a {\em Morse function} on $M$ if its critical points are nondegenerate, i.e., if the Hessian of $f$ at $x_c$ has only nonzero eigenvalues, so the critical points $x_c$ are isolated. Let us now consider a Hamiltonian dynamical system whose Hamiltonian is of the form
\beq
{\cal H} = \frac{1}{2}\sum_{i=1}^N \pi_i^2 + V(\varphi)~,
\label{H}
\eeq
where the $\varphi$'s and the $\pi$'s are, respectively, the coordinates and the conjugate momenta. The dynamics of such a system is defined in the 
$2N$-dimensional phase space spanned by the $\varphi$'s and the $\pi$'s, so that a natural choice would be to investigate the topology of the phase space by using the Hamiltonian ${\cal H}$ itself as a Morse function. However, for all the critical points of ${\cal H}$, $\pi_i = 0$ $\forall i$ holds, so that there is no loss of information in considering as our manifold $M$ the $N$-dimensional configuration space, and the potential
energy per particle ${\cal V}(\varphi) = V(\varphi)/N$ as our Morse function \cite{note_statmech}. Let us now consider the 
following family of submanifolds of $M$,
\beq
M_v = {\cal V}^{-1} (-\infty,v] = 
\{ \varphi \in M : {\cal V}(\varphi) \leq v\}~,
\eeq
i.e., each $M_v$ is the set $\{\varphi_i\}_{i=1}^N$ such that the potential energy per particle does not exceed a given value $v$. 
As $v$ is increased from $-\infty$ to $+\infty$, this
family covers successively the whole manifold $M$ \cite{note_limits}. All these manifolds have the same topology (homotopy type) until a critical level ${\cal V}^{-1}(v_c) = \partial M_{v_c}$ is crossed; here the topology of $M_v$ changes. More precisely, 
the manifolds $M_{v'}$ and $M_{v''}$, with $v' < v_c < v''$, 
can not be smoothly mapped onto each other if $v_c$ is a critical value of 
${\cal V}(\varphi)$.
A change in the topology of $M_v$ can only occur when $v$ passes through a 
critical value of ${\cal V}$. Thus in order to detect topological 
changes in $M_v$ we have to find the 
critical values of ${\cal V}$, which means solving the equations
\beq
\frac{\partial {\cal V}(\varphi)}{\partial \varphi_i} = 0~, 
\qquad i = 1,\ldots,N~.
\label{crit_eqs}
\eeq

Let us now apply this method to the case of the mean-field XY model. This model is particularly interesting four our purpose because the mean-field character of the interaction greatly simplifies the analysis \cite{method_note}, allowing an analytical treatment of the Eqs.\ (\ref{crit_eqs}); moreover, a projection of the configuration space onto a two-dimensional plane is possible. The mean-field XY model \cite{Antoni} describes a system of $N$ equally coupled planar classical rotators. It is defined by a Hamiltonian of the class (\ref{H}) where the potential energy is 
\beq
V(\varphi) = \frac{J}{2N}\sum_{i,j=1}^N 
\left[ 1 - \cos(\varphi_i - \varphi_j)\right] -h\sum_{i=1}^N \cos\varphi_i ~.
\label{V}
\eeq
Here $\varphi_i \in [0,2\pi]$ is the rotation angle of the $i$-th rotator and $h$ is an external field. Defining at each site $i$ a 
classical spin vector ${\bf s}_i = (\cos\varphi_i,\sin\varphi_i)$ the model 
describes a planar (XY) Heisenberg system with interactions of equal strength 
among all the spins. We consider only the ferromagnetic case $J >0$; 
for the sake of simplicity, we set $J=1$. The equilibrium statistical mechanics 
of this system is exactly described, in the thermodynamic limit, by mean-field 
theory \cite{Antoni}. In the limit $h \to 0^+$, the system has a continuous phase transition, with classical critical exponents, at $T_c = 1/2$, or $\varepsilon_c = 3/4$, 
where $\varepsilon = E/N$ is the energy per particle. 

We aim at showing that this phase transition has its foundation in a
basic topological change that occurs in the configuration space $M$ 
of the system. Let us remark that since ${\cal V}(\varphi)$ is bounded, $-h \leq {\cal V}(\varphi) \leq 1/2 + h^2/2$, the manifold is empty as long as $v< -h$, and when $v$ 
increases beyond  $1/2 + h^2/2$ no changes in its topology 
can occur so that the manifold $M_v$ remains the same for any $v > 1/2 + h^2/2$, and is an $N$-torus. To detect topological changes we have to solve Eqs.\ 
(\ref{crit_eqs}). To this end it is useful to define the  magnetization vector, i.e., the collective spin 
vector ${\bf m} = \frac{1}{N} \sum_{i=1}^N {\bf s}_i$, which as a 
function of the angles is given by 
\beq
{\bf m} = (m_x,m_y) = \left(\frac{1}{N}\sum_{i=1}^N \cos\varphi_i,
\frac{1}{N}\sum_{i=1}^N \sin\varphi_i  \right)~.
\label{m}
\eeq
Since, due to the mean-field character of the model, the potential energy
(\ref{V}) can be written as a function of ${\bf m}$ alone 
(remember that $J=1$), the potential energy per particle reads
\beq
{\cal V}(\varphi) = {\cal V}(m_x,m_y) = \frac{1}{2} (1 - m_x^2 - m_y^2) - h\, m_x~.
\label{V(m)}
\eeq
This allows us to write the Eqs. (\ref{crit_eqs}) in the form ($i = 1,\ldots,N$)
\beq
(m_x + h) \sin\varphi_i - m_y \cos\varphi_i = 0 ~ .
\label{crit_eqs_i}
\eeq
Now we can solve these equations and find
all the critical values of ${\cal V}$.
The solutions of Eqs.\ (\ref{crit_eqs_i}) can be grouped in three classes:

$(i)$ The minimal energy configuration $\varphi_i = 0 ~ 
\forall i$, with a critical value $v=v_0=-h$, which tends to 0 as $h \to 0^+$. In this case, $m_x^2 + m_y^2 = 1$.

$(ii)$ Configurations such that $m_y = 0, \sin\varphi_i = 0 ~
\forall i$. These are the configurations in which $\varphi_i$ equals either $0$ or $\pi$; i.e., we have again $\varphi_i = 0 ~ \forall i$, but also the $N$ configurations with $\varphi_k = \pi$ and $\varphi_i = 0 ~ \forall i \not = k$, then the $N(N-1)$ configurations with 2 angles equal to $\pi$ and all the others equal to 0, and so on, up to the configuration with $\varphi_i = \pi ~ \forall i$. The critical values corresponding to these critical points depend only on the number of $\pi$'s, $n_\pi$, so that $v(n_\pi) = \frac{1}{2}[1 - \frac{1}{N^2}(N - 2n_\pi)^2] - \frac{h}{N}(N - 2n_\pi)$. We see that the largest critical value is, for $N$ even, $v(n_\pi = N/2) = 1/2$ and that the number of critical points corresponding to it is ${\cal O}(2^N)$. 

$(iii)$ Configurations such that $m_x = -h$ and $m_y = 0$, which correspond to the critical value $v_c = 1/2 + h^2/2$, which tends to $1/2$ as $h \to 0^+$. The number of these configurations grows with $N$ not slower than $N!$ \cite{factorial_note}.

Configurations $(i)$ are the absolute minima of ${\cal V}$, $(iii)$ are the absolute maxima, and $(ii)$ are all the other stationary configurations of $\cal V$. 

Since for $v < v_0$ the manifold is empty, the topological change that
occurs at $v_0$ is the one corresponding to the ``birth'' of the manifold from the empty set; subsequently there are many topological changes at values $v(n_\pi) \in (v_0,1/2]$ till at $v_c$
there is a final topological change which corresponds to the ``completion''
of the manifold. We remark that the number of critical values in the interval $[v_0, 1/2]$ grows with $N$ and that eventually the set of these critical values becomes dense in the limit $N \to \infty$. However, the critical value $v_c$ remains isolated also in that limit.
We observe that considering a nonzero external field $h$ is necessary in order that $\cal V$ is a Morse function, because if $h = 0$ all the critical points of classes $(i)$ and $(ii)$ are degenerate, in which case topological changes do not necessarily occur \cite{Morse_theory}. This degeneracy is due to the 
$O(2)$-invariance of the potential energy in the absence of an external field. To be sure, for $h \not = 0$, $\cal V$ may not be a Morse function on the whole of $M$ either, but only on $M_v$ with $v < v_c$, because the critical points of class $(iii)$ may also be degenerate, so that $v_c$ does not necessarily correspond to a topological change. 

\begin{figure}
\vspace{0.25cm}
\centerline{\psfig{file=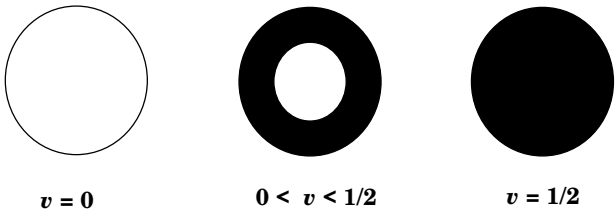,height=2.75cm}}
\vspace{0.25cm}
\caption{The sequence of topological changes undergone by the manifolds
$D_v$ with increasing $v$ in the limit $h \to 0^+$.
\label{fig_effective}}
\end{figure}

However, this difficulty could be dealt with by using that the potential energy can be written in terms of the collective
variables $m_x$ and $m_y$ --- see Eq. (\ref{V(m)}). This implies that we consider the system of $N$ spins projected onto the two-dimensional configuration space of the collective spin variables. According to the definition (\ref{m}) of ${\bf m}$, the accessible configuration space is now not the whole plane, but only the disk 
\beq
D = \{(m_x,m_y) : m_x^2 + m_y^2 \leq 1\}~.
\eeq
Thus we want to study the topology of the submanifolds 
\beq
D_v = \{ (m_x,m_y) \in D : {\cal V}(m_x,m_y) \leq v\}~.
\eeq
The sequence of topological transformations undergone by $D_v$ can now be
very simply determined in the limit $h \to 0^+$ 
(see Fig.\ \ref{fig_effective}). As long as $v < 0$, $D_v$ is the empty set. The first topological change occurs at $v = v_0 = 0$ where the manifold appears
as the circle $m_x^2 + m_y^2 = 1$, i.e., the boundary $\partial D$ of the
accessible region. Then as $v$ grows $D_v$ is given by the conditions
\beq
1 - 2v \leq m_x^2 + m_y^2 \leq 1~, 
\eeq 
i.e., it is the ring with a hole centered in $(0,0)$ (punctuated disk) 
comprised between the two circles
of radii $1$ and $\sqrt{2v}$, respectively. As $v$ continues to grow the hole shrinks and it is eventually completely filled as $v = v_c = 1/2$, where the second topological change occurs.  
In this coarse-grained two-dimensional description in $D$, all the topological changes that occur in $M$ between $v=0$ and $v=1/2$ disappear, and only the two topological changes corresponding to the extrema of $\cal V$, occurring at $v = v_0$ and $v = v_c$, survive. This strongly suggests that the topological change at $v_c$ should be present also in the full $N$-dimensional configuration space, so that the degeneracies mentioned above for the critical points of class $(iii)$ should not prevent a topological change.

Now we want to argue that the topological change occurring at 
$v_c$ is related to the thermodynamic phase transition of the mean-field XY model. Since the 
Hamiltonian is of the standard form (\ref{H}), the temperature $T$, the 
energy per particle $\varepsilon$ and the average potential energy per 
particle $u = \langle {\cal V} \rangle$ obey, 
in the thermodynamic limit, the following equation:
\beq
\varepsilon = \frac{T}{2} + u(T)~,
\eeq
where we have set Boltzmann's constant equal to 1. Substituting the values of the critical energy per particle 
$\varepsilon_c = 3/4$ and of the critical temperature $T_c = 1/2$ we get 
$u_c = u(T_c) = 1/2$, so that the critical value of the potential energy 
per particle $v_c$ where the last
topological change occurs equals the statistical-mechanical 
average value of the potential energy at the phase transition,
\beq
v_c = u_c~.
\label{vc=uc}
\eeq
Thus although a topological change in $M$ occurs at any $N$, and $v_c$ is {\em independent} of $N$, 
there is a connection of such a topological change and a thermodynamic phase 
transition {\em only} in the limit $N\to\infty$, $h \to 0^+$, when indeed thermodynamic phase transitions can be defined. 
A similar kind of difference, as here between topological changes in 
mathematics (for all $N$) and phase transitions in physics (for $N\to\infty$ 
only), also occurs in other contexts in statistical mechanics, e.g.\ 
in nonequilibrium stationary states \cite{CohenRondoni}.

The relevance of topologic concepts for phase transition theory
had already been proved, in the case of the two-dimensional Ising model
 \cite{Rasetti}, 
though in a rather abstract context: in that case the phase transition was
related to a jump in the Atiyah index of a suitable vector bundle.
However, Eq.\ (\ref{vc=uc}) strongly supports --- albeit for a special model --- the conjecture put forward in Ref.\ \cite{cccp}, 
i.e., that also the topology
of the configuration space changes in correspondence with a thermodynamic
phase transition. 

Since {\em not all} topological changes 
correspond to phase transitions, those that 
do correspond, remain to be determined to make the 
conjecture of Ref.\ \cite{cccp} more precise. 
In this context, we consider one example where there are topological changes very similar to the ones of our model but no phase 
transitions, i.e., the one-dimensional XY model with nearest-neighbor interactions, whose Hamiltonian is of the class (\ref{H}) with interaction potential
\beq
V (\varphi) = \frac{1}{4}\sum_{i=1}^N \left[ 1 - 
\cos(\varphi_{i+1} - \varphi_i) \right] -h\sum_{i=1}^N \cos\varphi_i ~.
\label{V_nn}
\eeq
In this case the configuration space $M$ is still an $N$-torus, and using again the specific interaction energy ${\cal V} = V/N$ as a Morse function we can prove that also here there are many topological changes in the submanifolds 
$M_v$ as $v$ is varied in the interval $[0,1/2]$ (after taking $h \to 0^+$). The critical points are all of the type $\varphi_j = \varphi_k = \varphi_l = \ldots = \pi$, $\varphi_i = 0 ~ \forall i \not = j,k,l,\ldots$; however, at variance with the mean-field XY model, it is no longer the number of $\pi$'s that determines the value of $\cal V$ at the critical point, but rather the number of domain walls, $n_d$, i.e., the number of boundaries between ``islands'' of $\pi$'s and ``islands'' of $0$'s: $v(n_d) = n_d/2N$. Since $n_d \in [0,N]$, the critical values lie in the same interval as in the case of the mean-field XY model. But now the maximum critical value $v = 1/2$, instead of corresponding to a huge number of critical points, which rapidly grows with $N$, corresponds to {\em only two} configurations with $N$ domain walls, which are $\varphi_{2k} = 0$, $\varphi_{2k + 1} = \pi$, with $k = 1,\ldots,N/2$, and the reversed one.

Thus this example suggests the conjecture that a topological change in the configuration space submanifolds $M_v$ occurring at a critical value $v_c$ is associated with a phase transition in the thermodynamic limit if the number of critical points corresponding to the critical value $v_c$ is sufficiently rapidly growing with $N$. On the basis of the behavior of the mean-field XY model we expect that such a growth should be at least exponential. Furthermore, a relevant feature appears to be that $v_c$ remains an isolated critical value also in the limit $N \to \infty$: in the mean-field XY model this holds only if the thermodynamic limit is taken {\em before} the $h \to 0^+$ limit: this appears as a topological counterpart of the 
non-commutativity of the limits $h \to 0^+$ and $N \to \infty$ in order to get a phase transition in statistical mechanics.

We conclude with some comments. 
The sequence of topological changes occurring with growing $\cal V$
makes the configuration space larger and larger, till at $v_c$ the whole configuration space becomes fully accessible to the system through the last topological change. From a physical point of view, this corresponds to the appearance of more and more disordered configurations as $T$ grows, which ultimately lead to the phase transition at $T_c$. We remark that the connection between the topology of the configuration space and the physics of continuous phase transitions made here via the potential energy, in particular Eq.\ (\ref{vc=uc}), only makes sense in the thermodynamic limit, where the potential energy per particle $u(T)$ is well-defined since its fluctuations vanish then at least as $1/\sqrt{N}$ \protect\cite{mf_note}.

Since a notion of universality arises quite naturally in a topological framework, it is tempting to think that
universal quantities like critical exponents might have in general a topological
counterpart. 
Finally, the fact that the topological changes appear at any $N$
opens a new possibility to study transitional phenomena in {\em finite}
systems, like atomic clusters, nuclei, polymers and proteins, or other 
biological systems. 

We thank R.\ Livi and M.\ Rasetti for useful discussions. 
LC gratefully acknowledges the Rockefeller University in New York 
for its kind hospitality and for partial financial support. 
EGDC is indebted to the US Department of Energy for support under 
grant DE-FG02-88-ER13847, and to the Politecnico di Torino and INFM -- UdR Torino Politecnico for hospitality and partial financial support.  

\end{multicols}

\begin{references}
\bibitem{note_statmech} This is similar to what happens in the statistical mechanics of Hamiltonian systems of the form (\protect\ref{H}), where the momenta can be integrated out and the statistical measure can be defined on the configuration space alone.
\bibitem{note_limits} If $\cal V$ is bounded below, as many physical potentials are, then the manifolds $M_v$ are empty as long as $v < \min({\cal V})$.
\bibitem{factorial_note} Such an estimate follows from the 
construction of a Morse function on an $N$-torus whose critical points are the configurations with ${\bf m} = {\bf 0}$, so that the number of such configurations can not be smaller than the sum of the Betti numbers of an 
$N$-torus, which grows at least as $N!$.
\bibitem{mf_note} This holds for our mean-field model,
since fluctuations vanish. In the case of a real continuous (critical) phase transition the non-trivial role of
fluctuations may complicate the present picture.
\end{references}
\end{document}